\documentclass[12pt]{article}
\usepackage{amssymb}
\usepackage{amsmath}
\usepackage{amsthm}
\usepackage{hyperref}
\usepackage{cite}

\begin{document}

%************************** Text Begins here ******************************

%  Greek letters

\def\a{\alpha}
\def\b{\beta}
\def\d{\delta}
\def\e{\epsilon}
\def\g{\gamma}
\def\h{\mathfrak{h}}
\def\k{\kappa}
\def\l{\lambda}
\def\o{\omega}
\def\p{\wp}
\def\r{\rho}
\def\t{\tau}
\def\s{\sigma}
\def\z{\zeta}
\def\x{\xi}
\def\V={{{\bf\rm{V}}}}
 \def\A{{\cal{A}}}
 \def\B{{\cal{B}}}
 \def\C{{\cal{C}}}
 \def\D{{\cal{D}}}
\def\G{\Gamma}
\def\K{{\cal{K}}}
\def\O{\Omega}
\def\R{\bar{R}}
\def\T{{\cal{T}}}
\def\L{\Lambda}
\def\f{E_{\tau,\eta}(sl_2)}
\def\E{E_{\tau,\eta}(sl_n)}
\def\Zb{\mathbb{Z}}
\def\Cb{\mathbb{C}}
\def\ve{\varepsilon}
\def\tr{{\rm tr}}
\def\E{{\rm e}}
\def\i{{\rm i}}

\def\R{\overline{R}}
% Shorthands for \begin{equation} and the like

\def\beq{\begin{equation}}
\def\eeq{\end{equation}}
\def\bea{\begin{eqnarray}}
\def\eea{\end{eqnarray}}
\def\ba{\begin{array}}
\def\ea{\end{array}}
\def\no{\nonumber}
\def\le{\langle}
\def\re{\rangle}
\def\lt{\left}
\def\rt{\right}
\def\id{\mathbb{I}}
\newtheorem{Theorem}{Theorem}
\newtheorem{Definition}{Definition}
\newtheorem{Proposition}{Proposition}
\newtheorem{Lemma}{Lemma}
\newtheorem{Corollary}{Corollary}
\renewcommand{\theremark}{\thesection.\arabic{remark}}

\renewcommand{\thefootnote}{\fnsymbol{footnote}}
 \setcounter{footnote}{0}

\newfont{\elevenmib}{cmmib10 scaled\magstep1}
\newcommand{\preprint}{
   \begin{flushleft}
     %\elevenmib Yukawa\, Institute\, Kyoto\\
   \end{flushleft}\vspace{-1.3cm}
   \begin{flushright}\normalsize
  % \sf  YITP-03-53\\
   %  {\tt hep-th/yymmnnn} \\
   %July 2006
   \end{flushright}}
\newcommand{\Title}[1]{{\baselineskip=26pt
   \begin{center} \Large \bf #1 \\ \ \\ \end{center}}}
\newcommand{\Author}{\begin{center}
   \large \bf

Xiaotian Xu\textsuperscript{1,2,3},
Pei Sun\textsuperscript{2,3,4*},
Xin Zhang\textsuperscript{5\dag},
Junpeng Cao\textsuperscript{2,5},
Tao Yang\textsuperscript{1,2,3},

\end{center}}
\newcommand{\Address}{

\begin{center}
{\bf 1} Institute of Modern Physics, Northwest University, Xi'an 710127, China
\\
{\bf 2} Peng Huanwu Center for Fundamental Theory, Xi'an 710127, China
\\
{\bf 3} Shaanxi Key Laboratory for Theoretical Physics Frontiers, Xi'an 710127, China
\\
{\bf 4} School of Physics, Northwest University, Xi'an 710127, China
\\
{\bf 5} Beijing National Laboratory for Condensed Matter Physics, Institute of Physics, Chinese Academy of Sciences, Beijing 100190, China
\\
% TODO: provide email address of corresponding author
*sunpeiphy@163.com
\dag xinzhang@iphy.ac.cn
\end{center}
}
\newcommand{\Accepted}[1]{\begin{center}
   {\large \sf #1}\\ \vspace{1mm}{\small \sf Accepted for Publication}
   \end{center}}

\preprint
\thispagestyle{empty}
\bigskip\bigskip\bigskip
\Title{Exact solution of the Izergin-Korepin Gaudin model with the generic open boundaries
} \Author

\Address
\vspace{1cm}

\begin{abstract}
We study the Izergin-Korepin Gaudin models with both periodic and open integrable boundary conditions, which describe quantum systems exhibiting novel long-range interactions. Using the Bethe ansatz approach, we derive the eigenvalues of the Gaudin operators and the corresponding Bethe ansatz equations.

\vspace{1truecm} \noindent {\it PACS:} 03.65.Fd; 04.20.Jb; 05.30.-d; 75.10.Jm

\end{abstract}

\newpage

\setcounter{footnote}{0}
\renewcommand{\thefootnote}{\arabic{footnote}}

\baselineskip=20pt

\section{Introduction}
\setcounter{equation}{0}
The Gaudin model \cite{Gau76} describes an important class of one-dimensional many-body systems with long-range interactions and has widespread applications in various research fields, such as condensed matter physics and high-energy physics. For example, they are relevant in the simplified BCS theory for small metallic particles \cite{Amico:2001qn, Dukelsky:2004re} and in the Seiberg-Witten theory of super-symmetric gauge theory \cite{Sei94, Bra99}. In addition, Gaudin models provide powerful tools for constructing integral representations of solutions to the Knizhnik-Zamolodchikov equations \cite{KZ, Hik95, H.M93, Feign94}.

The Gaudin operators with integrable boundary conditions can be constructed through a quasi-classical expansion of the inhomogeneous transfer matrix of quantum integrable models \cite{Hikami92, Hik95, Lor02}. Within this framework, one can diagonalize the Gaudin operators once the exact solutions of the corresponding transfer matrix are derived. Following Gaudin's pioneering work, various integrable Gaudin models have been constructed and solved exactly \cite{Skl87,Skl96,Kulish:2001,Dukelsky:2001fe,Lima-Santos:2001,Yang04a,Yan04,Yan05,CiriloAntonio:2013,Hao:2014wza,Mukhin:2014,Manojlovic2017,Manojlovic:2,Crampe:2017}.
Among these integrable models, the most well-studied ones are those with $U(1)$ symmetry, where the conventional Bethe ansatz approaches works well.

On the other hand, advancements in several analytical methods-such as the generalized Bethe ansatz method \cite{Cao03,Belliard:2013,Zhang21}, the functional $T$-$Q$ relation \cite{Nepomechie03a,Nepomechie03b,Yang:2005}, and the off-diagonal Bethe ansatz method \cite{ODBA,Cao1a,Cao1b,Cao1d}-have enabled us to solve non-trivial integrable models that lack $U(1)$ symmetry \cite{ODBA,Li14,Zha14,Hao14}. These progress motivate us to explore novel Gaudin models and analyze their exact solutions.

In this paper, we study the Izergin-Korepin (IK) Gaudin model with periodic and open boundary conditions. The IK model plays a fundamental role in the study of non-$A$-type integrable models \cite{IK81}. Exact solutions of the IK model with periodic and generic open boundaries have been constructed using the algebraic Bethe ansatz \cite{Tarasov88} and the off-diagonal Bethe ansatz \cite{ODBA,Hao14}, respectively. The exactly solvable IK Gaudin model is constructed by following the standard approach, i.e., by proceeding with a quasi-classical expansion of the corresponding inhomogeneous transfer matrix \cite{Lor02,Hao:2014wza}. After some analytic calculations, we obtain the exact spectrum of the IK Gaudin model, which is parameterized by the solutions of the Bethe ansatz equations (BAEs).

The paper is organized as follows. In Section 2, we introduce the IK model with periodic boundary conditions and its exact solutions. Section 3 focuses on the construction of the IK Gaudin operators under periodic boundary conditions, and provides their solutions-including their eigenvalues and the corresponding BAEs. In Section 4, we present the IK model with open boundaries and demonstrate its exact solutions. Section 5 focuses on constructing the IK Gaudin model with open boundaries. In Section 6, we derive the eigenvalues of the open Gaudin operators and give their corresponding BAEs. The last section provides a summary of our results.

\section{The IK model with periodic boundaries}
\setcounter{equation}{0}
\subsection{Integrability of periodic IK model}

The $R$-matrix of the IK model \cite{IK81}, associated with the simplest twisted affine algebra $A_2^{(2)}$, is given by
\renewcommand{\arraystretch}{1.2}
\begin{align}
R(u)=\left(
\begin{array}{ccc|ccc|ccc}
 c(u) & 0 & 0 & 0 & 0 & 0 & 0 & 0 & 0 \\
 0 & b(u) & 0 & e(u) & 0 & 0 & 0 & 0 & 0 \\
 0 & 0 & d(u) & 0 & g(u) & 0 & f(u) & 0 & 0 \\
 \hline
 0 & \bar e(u) & 0 & b(u) & 0 & 0 & 0 & 0 & 0 \\
 0 & 0 & \bar g(u) & 0 & a(u) & 0 & g(u) & 0 & 0 \\
 0 & 0 & 0 & 0 & 0 & b(u) & 0 & e(u) & 0 \\
 \hline
 0 & 0 & \bar f(u) & 0 & \bar g(u) & 0 & d(u) & 0 & 0 \\
 0 & 0 & 0 & 0 & 0 & \bar e(u) & 0 & b(u) & 0 \\
 0 & 0 & 0 & 0 & 0 & 0 & 0 & 0 & c(u) \\
\end{array}
\right).\label{R;Matrix}
\end{align}
The expressions for the functions in Eq. (\ref{R;Matrix}) are
\begin{align}
\begin{aligned}
&a(u)=\sinh(u-3\eta)
-\sinh 5\eta+\sinh 3\eta+\sinh\eta,\quad
b(u)=\sinh(u-3\eta)
+\sinh3\eta,\\[6pt]
&c(u)=\sinh(u-5\eta)+\sinh\eta,\quad d(u)=\sinh(u-\eta)+\sinh\eta,\\
&e(u)=-2\E^{-\frac{u}{2}}\sinh2\eta\cosh(\tfrac{u}{2}-3\eta),\quad \bar{e}(u)=-2\E^{\frac{u}{2}}\sinh2\eta\cosh(\tfrac{u}{2}-3\eta),\\
&f(u)=-2 \E^{-u+2\eta}\sinh\eta\sinh2\eta-\E^{-\eta}\sinh4\eta,\\[6pt] &\bar{f}(u)=2 \E^{u-2\eta}\sinh\eta\sinh2\eta-\E^{\eta}\sinh4\eta,\\[6pt]
&g(u)=2\E^{-\frac{u}{2}+2\eta}\sinh\tfrac{u}{2}\sinh 2\eta,\quad \bar{g}(u)=-2\E^{\frac{u}{2}-2\eta}\sinh\tfrac{u}{2}\sinh 2\eta.
\label{R-element-2}
\end{aligned}
\end{align}

The $R$-matrix in (\ref{R;Matrix}) satisfies the quantum Yang-Baxter equation (QYBE) \cite{Korepin;book}
\bea
R_{1,2}(u_1-u_2)R_{1,3}(u_1-u_3)R_{2,3}(u_2-u_3)
=R_{2,3}(u_2-u_3)R_{1,3}(u_1-u_3)R_{1,2}(u_1-u_2),\label{QYB}\eea
and possesses the following properties:
\begin{align}
&\mbox{Initial
condition}:\quad R_{1,2}(0)= \kappa P_{1,2},\quad \kappa=\sinh \eta-\sinh 5\eta,\label{Int-R}\\
&\mbox{Unitarity
relation}:\quad R_{1,2}(u)R_{2,1}(-u)= c(u)c(-u)\,\times {\id}_{1,2},\label{Unitarity}\\
&\mbox{Crossing
relation}:\quad R_{1,2}(u)=V_1R_{1,2}^{t_2}(-u+6\eta+\i\pi)V^{-1}_1,\nonumber\\
&\hspace{3.5cm}\quad V=\left(\begin{array}{ccc}
     &  & -\E^{-\eta}\\
     & 1 & \\
   -\E^{\eta}  &  &
\end{array}\right),
\label{crosing}\\
&\mbox{Quasi-classical property}:\quad R(u)|_{\eta\rightarrow0}=\sinh u\times {\id},\label{cond-1}
\end{align}
where $\id$ is the identity matrix, $P_{1,2}$ is
the permutation operator, $R_{2,1}(u)=P_{1,2}R_{1,2}(u)P_{1,2}$, and the superscript $t_i$ indicates the transposition in the $i$-th space.

In the framework of the algebraic Bethe ansatz method \cite{Korepin;book}, one can construct the transfer matrix
\begin{align}
t^{(p)}(u)=\mathrm{tr}_0\{R_{0,N}(u-\theta_N)R_{0,N-1}(u-\theta_{N-1})\cdots
R_{0,1}(u-\theta_1)\},
\end{align}
where $\{\theta_1,\ldots,\theta_N\}$ is a set of inhomogeneous parameters. Here and below, the superscript $(p)$ indicates that the system is under the periodic boundary conditions.

By using the QYBE (\ref{QYB}) repeatedly,
one can demonstrate that the transfer matrices with
different spectral parameters commute with each other \cite{Skl88} :
\bea
[t^{(p)}(u),\,\,t^{(p)}(v)]=0\label{com-1}.
\eea
The transfer matrix $t^{(p)}(u)$ acts as the generating functional of the conserved quantities of the system and the integrability of the system is thus proved.

\subsection{Exact solutions of periodic IK model}
Introduce some functions
\begin{align}
\Tilde{\mathbf{a}}(u)&=\prod_{l=1}^Nc(u-\theta_l),\\
\Tilde{\mathbf{d}}(u)&=\prod_{l=1}^Nd(u-\theta_l),\\
\Tilde{\mathbf{b}}(u)&=\prod_{l=1}^Nb(u-\theta_l).
\end{align}
With the help of the conventional Bethe ansatz method, the eigenvalues of the transfer matrix $t^{(p)}(u)$ can be parameterize by the following $T$-$Q$ relation \cite{Tarasov88,Martins95}
\begin{align}
\L^{(p)}(u)=\Tilde{\mathbf{a}}(u)\frac{\widetilde{Q}(u+4\eta)}{\widetilde{Q}(u)}+\Tilde{\mathbf{d}}(u)\frac{\widetilde{Q}(u-6\eta+\i\pi)}{\widetilde{Q}(u-2\eta+\i\pi)}+\Tilde{\mathbf{b}}(u)\frac{\widetilde{Q}(u-4\eta)\widetilde{Q}(u+2\eta+\i\pi)}{\widetilde{Q}(u-2\eta+\i\pi)\widetilde{Q}(u)},
\label{T-Q-p}
\end{align}
where
\bea
\widetilde{Q}(u)=\prod_{j=1}^M\sinh\Big(\frac{u-\l_j-2\eta}{2}\Big),\label{Qp}
\eea
and $M=0,1,\ldots,N$.
The Bethe roots $\{\l_1,\ldots,\l_M\}$ in Eq. (\ref{Qp}) satisfy the following Bethe ansatz equations (BAEs)
\bea
&&\prod_{l=1}^N\frac{\sinh\Big(\frac{\l_j-\theta_l-2\eta}{2}\Big)}
{\sinh\Big(\frac{\l_j-\theta_l+2\eta}{2}\Big)}=-\frac{\widetilde{Q}(\l_j-2\eta)\widetilde{Q}(\l_j+4\eta+\i\pi)}{\widetilde{Q}(\l_j+6\eta)\widetilde{Q}(\l_j+\i\pi)},\quad j=1,\ldots,M.\label{BAE-p}
\eea

\section{IK Gaudin model with periodic boundaries and its exact solutions}\label{Gaudin;PBC}
\setcounter{equation}{0}

\subsection{Construction of Gaudin operators}\label{Sec;Gaudin;Periodic}

The IK Gaudin operators $\{H_1^{(p)},\ldots,H_N^{(p)}\}$ with periodic boundary conditions can be constructed by expanding the transfer matrix at the point $u=\theta_j$ and around $\eta=0$ as follows
\begin{align}
&t^{(p)}(\theta_j)=\kappa\,\mathsf{t}_0^{(p)}(\theta_j)\Big(\id+\eta H_j^{(p)}+\cdots\Big), \quad j=1,\dots, N,\label{t-p-1}\\[6pt]
&H_j^{(p)}=\lt.\frac{\partial\ln( t^{(p)}(\theta_j)/\kappa)}{\partial\eta}\rt|_{\eta=0}.
\label{exp-p}
\end{align}
From Refs. \cite{Cao1d,ODBA}, we know that
\begin{align}
t^{(p)}(\theta_j)=&\,\kappa R_{j,j-1}(\theta_j-\theta_{j-1})\cdots R_{j,1}(\theta_j-\theta_1)R_{j,N}(\theta_j-\theta_N)\cdots R_{j,j+1}(\theta_j-\theta_{j+1}).
\end{align}
The quasi-classical property of the $R$-matrix shown in Eq. (\ref{cond-1}) allows us to introduce the corresponding classical $r$-matrix $r(u)$
\begin{align}
R(u)=&\sinh u\times{\id}+\eta\, r(u)+O(\eta^2),\qquad{\rm when}~~ \eta\rightarrow 0,\no\\
r(u)=&\left.\frac{\partial R(u) }{\partial\eta}\right|_{\eta=0}.\label{ope-r}
\end{align}

The matrix representation of $r(u)$ is
\small{\begin{align}
r(u)=\left(
\begin{array}{ccc|ccc|ccc}
 c'(u) & 0 & 0 & 0 & 0 & 0 & 0 & 0 & 0 \\
 0 & b'(u) & 0 & e'(u) & 0 & 0 & 0 & 0 & 0 \\
 0 & 0 & d'(u) & 0 & g'(u) & 0 & f'(u) & 0 & 0 \\
 \hline
 0 & \bar e'(u) & 0 & b'(u) & 0 & 0 & 0 & 0 & 0 \\
 0 & 0 & \bar g'(u) & 0 & a'(u) & 0 & g'(u) & 0 & 0 \\
 0 & 0 & 0 & 0 & 0 & b'(u) & 0 & e'(u) & 0 \\
 \hline
 0 & 0 & \bar f'(u) & 0 & \bar g'(u) & 0 & d'(u) & 0 & 0 \\
 0 & 0 & 0 & 0 & 0 & \bar e'(u) & 0 & b'(u) & 0 \\
 0 & 0 & 0 & 0 & 0 & 0 & 0 & 0 & c'(u) \\
\end{array}
\right)\label{R-3-1}.
\end{align}}
The non-zero entries in Eq. (\ref{R-3-1}) read
\bea
&&a'(u)=-3\cosh u-1,\quad b'(u)=3-3\cosh u,\quad c'(u)=1-5\cosh u,\no\\
&&d'(u)=1-\cosh u,\quad e'(u)=-4\E^{-\frac{u}{2}}\cosh\tfrac{u}{2},\quad \bar e'(u)=-4\E^{\frac{u}{2}}\cosh\tfrac{u}{2},\no\\
&&f'(u)=-4,\quad \bar f'(u)=-4, \quad g'(u)=4\E^{-\frac{u}{2}}\sinh\tfrac{u}{2},\quad \bar g'(u)=-4\E^{\frac{u}{2}}\sinh\tfrac{u}{2}.
\eea
With the help of Eq. (\ref{cond-1}), we can obtain the expression for $\mathsf{t}_0^{(p)}(\theta_j)$ and the corresponding Gaudin operators $H_j^{(p)}$
\begin{align}
\mathsf{t}_0^{(p)}(\theta_j)&=\prod_{l\neq j}^{N}\sinh(\theta_j-\theta_l)\times{\id},\label{Ope-p}\\
H_j^{(p)}&=\sum_{l\neq j}^{N}\Gamma_{j,l}(\theta_j,\theta_l),\quad \Gamma_{j,l}(\theta_j,\theta_l)=\frac{r_{j,l}(\theta_j-\theta_l)}{\sinh(\theta_j-\theta_l)}\label{new-1}.
\end{align}

Here $\Gamma_{j,l}(\theta_j,\theta_l)$ describes a long-range two-site interactions between sites $j$ and $l$ (with $l\neq j)$, which only depends on the inhomogeneous parameters $\theta_j$ and $\theta_l$.
We can use the spin-1 operators $S^{\alpha},\,\alpha=\pm,z$ to expand the operator $r_{j,l}(u)$ as follows
\begin{align}
r_{j,l}(u)=&-(3\cosh u+1) \id_{j,l}+4 [(S_j^z)^2+(S_l^z)^2]-6(S_j^z)^2(S_l^z)^2-2\cosh u\, S_j^z S_l^z\no\\
&-(S_j^+)^2(S_l^-)^2-(S_j^-)^2(S_l^+)^2-2 \E^{-\frac{u}{2}} \cosh (\tfrac{u}{2})(S_j^zS_j^+S_l^-S_l^z+S_j^+S_j^zS_l^zS_l^-)\no\\
&-2\E^{-\frac{u}{2}} \sinh (\tfrac{u}{2})(S_j^zS_j^+S_l^zS_l^-+S_j^+S_j^zS_l^-S_l^z)-2\E^{\frac{u}{2}} \cosh (\tfrac{u}{2})(S_j^-S_j^zS_l^zS_l^+\no\\
&+S_j^zS_j^-S_l^+S_l^z)+2\E^{\frac{u}{2}} \sinh (\tfrac{u}{2})(S_j^-S_j^zS_l^+S_l^z+S_j^zS_j^-S_l^zS_l^+).\label{operator;r}
\end{align}

One see that the operator
$r(u)$ is Hermitian when
$\i u\in\mathbb{R}$. As a consequence, a family of Hermitian operators
$\{\i H_j^{(p)}\}$ can be obtained if the parameters $\{\theta_j\}$ all lie on the imaginary axis.

Based on the expansion of $t^{(p)}(\theta_j)$ with respect to $\eta$ (\ref{t-p-1}) and the commutation relation of the transfer matrix with different spectral parameters (\ref{com-1}), we can prove that $\{H_j^{(p)}\}$ mutually commute. The proof is as follows.
\begin{proof}
For convenience, we omit the symbol $(p)$ from $\mathsf{t}_0^{(p)}(\theta_j)$ and $H_j^{(p)}$ in the proof. From Eq. (\ref{Ope-p}), we see that $\mathsf{t}_0(\theta_j)$ is proportional to the identity matrix and commutes with any operators. The commutation relation $[t(\theta_j),\,t(\theta_l)]=0$ then leads to
\begin{align}
   &\quad[\id+\eta H_j+\eta^2 H^{(2)}_j+\cdots,\,\id+\eta H_l+\eta^2 H^{(2)}_l+\cdots]\no\\
   &=[\mathbb{I},\,\id]+\eta\Big\{[\id,\,H_l]+[H_j,\,\id]\Big\}\no\\
&\quad+\eta^2\Big\{[H_j^{(2)},\,\id]+[\id,\,H_l^{(2)}]+[H_j,\,H_l]\Big\}+\cdots\no\\
&=\eta^2[H_j,\,H_l]+\eta^3(\cdots)+\cdots=0,\label{comu-p-3}
\end{align}
Since $\eta$ is arbitrary, the coefficients of each power of $\eta$ in (\ref{comu-p-3}) must be zero. On examining the $\eta^2$ term specifically, this yields the following equation
\begin{align}
[H_j,\,H_l]=0.
\end{align}
\end{proof}
The aforementioned proof is also valid for the open system. It should be remarked that we require $ \mathsf{t}_0^{(p)}(\theta_j)=\lim\limits_{\eta \to 0} t(\theta_j)/\kappa$ to be proportional to the identity operator. This condition is automatically satisfied in the periodic system. However, for the open system, additional constraints on the model parameters are mandated so that the condition holds (see Eq. (\ref{cons-open})).\subsection{Exact solutions of Gaudin operators}

The Gaudin operator $H_j^{(p)}$ is exactly solvable.
We can derive the eigenvalues of the Gaudin operators directly from the $T$-$Q$ relation of the transfer matrix (\ref{T-Q-p}). The Bethe roots in the $T$-$Q$ relation are also functions of the parameter $\eta$. The Bethe roots $\{\lambda_j|_{j=1,\dots,M}\}$ can be expanded in terms of $\eta$ as follows:
\bea
\lambda_j=\mu_j+\eta\nu_j+O(\eta^2).
\eea
By setting $u=\theta_j$ in the $T$-$Q$ relation (\ref{T-Q-p}) and taking the first derivative of $\ln\Lambda^{(p)}(\theta_j)$ with respect to $\eta$ at $\eta=0$, we finally can get the eigenvalue of the periodic IK Gaudin operator $E_j^{(p)}$, which read
\begin{align}
E_j^{(p)}&=\lt.\frac{\partial \ln(\Lambda^{(p)}(\theta_j)/\kappa)}{\partial\eta}\rt|_{\eta=0}\no\\[6pt]
&=\sum_{k\neq j}^N\frac{1-5\cosh(\theta_j-\theta_k)}{\sinh(\theta_j-\theta_k)}+2\sum_{k=1}^M\coth\Big(\frac{\theta_j-\mu_k}{2}\Big)\label{ej-p}.
\end{align}
The Bethe roots $\{\mu_j\}$ in Eq. (\ref{ej-p}) should satisfy the following BAEs
\bea
&&\sum_{l\neq j}^M\lt[-2\coth\Big(\frac{\mu_j-\mu_l}{2}\Big)+\tanh\Big(\frac{\mu_j-\mu_l}{2}\Big)\rt]\no\\
&&+\sum_{l=1}^N\coth\Big(\frac{\mu_j-\theta_l}{2}\Big)=0,\qquad j=1,\ldots,M.\label{bae-p}
\eea
\begin{table}[htbp]
\centering
\begin{tabular}{|c|c|c|r|c|}
\hline
 $\mu_1$& $\mu_2$ & $\mu_3$ & $\i E_2^{(p)}$~~ & $d$\\
 \hline
0.0000+0.4353$\i$ & 0.9196$-$1.5234$\i$ & $-$0.9196$-$1.5234$\i$ & $-$12.9430 & 1\\
0.0000$-$1.5502$\i$ & 0.0000+0.4521$\i$ & -- &$-$12.7628 &2\\
$-$1.3935+1.5603$\i$ & 0.0000+0.5509$\i$ & 1.3935+1.5603$\i$ & $-$12.5338 & 1\\
0.0000+0.5733$\i$ & 0.0000+1.5294$\i$ & -- &$-$12.3989  &2\\
0.0000+0.5056$\i$ & -- & -- & $-$12.0349 &2\\
0.0000$-$0.1551$\i$ & 0.0000+0.5085$\i$ & 0.0000$-$2.9649$\i$ & $-$0.1070 & 1\\
0.0000$-$1.2447$\i$ & 0.0000+3.3197$\i$ & 0.0000+1.5966$\i$ & 0.1198 &1\\
0.0000+1.5035$\i$ & 0.0000$-$1.1513$\i$ & -- &0.1204 &2\\
2.4667$-$2.9649$\i$ & $-$2.4667$-$2.9649$\i$ & 0.0000$-$2.9650$\i$ & 0.1354 & 1 \\
1.1525$-$2.9649$\i$ & $-$1.1525$-$2.9649$\i$ &  -- & 0.1358 &2 \\
0.0000$-$2.9650$\i$ & -- & -- & 0.1373  &2\\
-- & -- & -- & 0.1407 &2\\
0.0000$-$0.1522$\i$  & -- & -- & 12.0709 & 2\\
0.0000$-$0.2218$\i$ & 0.0000$-$1.1745$\i$ & -- & 12.4479 &2 \\
1.3948$-$1.2064$\i$ & 0.0000$-$0.1987$\i$ & $-$1.3948$-$1.2064$\i$ & 12.5850 & 1\\
0.0000+1.9031$\i$ & 0.0000$-$0.0970$\i$ &  --&12.7716 &2\\
0.0000$-$0.0796$\i$ & $-$0.9210+1.8756$\i$ & 0.9210+1.8756$\i$ & 12.9496 & 1 \\
 \hline
\end{tabular}
\caption{Numeric solutions of Eq. (\ref{bae-p}) with $\{\theta_1,\theta_2,\theta_3\}=\{-0.40\i,0.18 \i,0.75\i\}$. The eigenvalue of $\i H_2^{(p)}$ given by Eq. (\ref{ej-p}) matches the exact diagonalization results. Here $d$ represents the degeneracy of the energy level. }
\label{tab1}
\end{table}

From Eq. (\ref{operator;r}), we observe that the total $z$-component spin operator $\sum_j S_j^z$ commutes with the Gaudin operator $H_j^{(p)}$ and these operators share the same eigenstates. The integer $M$ now is a conserved charge representing the total number of spinons. When $M=0$, the energy is $\sum_k\frac{1-5\cosh(\theta_j-\theta_k)}{\sinh(\theta_j-\theta_k)}$, which corresponds to the vacuum state
$|1\rangle^{\otimes N}$.

Due to the $\mathbb{Z}_2$ symmetry, all eigenstates in the $M \neq N$ sectors exhibit degeneracy. Therefore, we only need to solve the BAEs (\ref{bae-p}) with $0\leq M\leq N$.
Some numerical solutions of Eq. (\ref{bae-p}) for small-scale systems are presented in Table \ref{tab1}.

\section{The IK model with open boundaries}\label{TQ;Open}
\setcounter{equation}{0}
\subsection{Integrability of open IK model}

For an integrable system with open boundaries, in addition to the $R$-matrix, we also require the boundary$-$related $K$-matrices \cite{Skl88}.
In this paper, we consider the type II non-diagonal $K$-matrices in Ref. \cite{Kim}
\begin{align}
K^{-}(u)&=\left(
          \begin{array}{ccc}
            1+2\E^{-u-\epsilon}\sinh\eta & 0 & 2\E^{-\epsilon+\sigma}\sinh u \\
            0 & 1-2\E^{-\epsilon}\sinh(u-\eta) & 0 \\
            2\E^{-\epsilon-\sigma}\sinh u & 0 & 1+2\E^{u-\epsilon}\sinh\eta \\
          \end{array}
        \right),\label{Kf}\\[6pt]
K^{+}(u)&={\cal M}K^{-}(-u+6\eta+\i\pi)\Big|_{(\epsilon,\sigma)\rightarrow(\epsilon^{\prime},\sigma^{\prime})},\label{Kz}\\[6pt]
{\cal M}&={\rm diag}\Big\{\E^{2\eta},\,1,\,\E^{-2\eta}\Big\}.\label{M-matrix}
\end{align}
The matrices $K^-(u)$ and $K^+(u)$ satisfy the reflection equation (RE) and the dual RE respectively \cite{Lima99,Nep2002}, specifically as follows

\bea &&R_{1,2}(u_1-u_2)K^-_1(u_1)R_{2,1}(u_1+u_2)K^-_2(u_2)\no\\
 &&~~~~~~=
K^-_2(u_2)R_{1,2}(u_1+u_2)K^-_1(u_1)R_{2,1}(u_1-u_2),\label{RE-V}\\
&&R_{1,2}(u_2-u_1)K^+_1(u_1)\mathcal{M}_1^{-1}R_{2,1}(-u_1-u_2+12\eta)\mathcal{M}_1K^+_2(u_2)\no\\
&&~~~~~~= K^+_2(u_2)\mathcal{M}_2^{-1}R_{1,2}(-u_1-u_2+12\eta)\mathcal{M}_2K^+_1(u_1)R_{2,1}(u_2-u_1).
\label{DRE-V}\eea

Then the double-row transfer matrix of the IK model is constructed
\begin{align}
t(u)=&\,\tr_0\{K^+_0(u)R_{0,N}(u-\theta_N)R_{0,N-1}(u-\theta_{N-1})\cdots
R_{0,1}(u-\theta_1)\no\\
&\times K^-_0(u)R_{1,0}(u+\theta_1)R_{2,0}(u+\theta_{2})\cdots
R_{N,0}(u+\theta_N)\}.\label{trans}
\end{align}

With the help of QYBE (\ref{QYB}) and (dual) REs
(\ref{RE-V}) and (\ref{DRE-V}),
one can prove that the transfer matrices with
different spectral parameters commute with each other \cite{Skl88} :
\bea
[t(u),\,t(v)]=0.
\eea
This ensures the integrability of the system. The transfer matrix (\ref{trans}) indeed does depend on the inhomogeneous parameters $\{\theta_j\}$ and four free boundary parameters
$\{\e,\,\sigma,\,\e',\,\sigma'\}$.
\subsection{\texorpdfstring{Inhomogeneous $T$-$Q$ relation}{Inhomogeneous T-Q relation}}
In Refs. \cite{ODBA,Hao14}, the transfer matrix $t(u)$ defined in (\ref{trans}) has been exactly diagonalized via the off-diagonal Bethe ansatz approach. Let us recall the $T$-$Q$ relation.

First, introduce some functions
\begin{align}
\mathbf{a}(u)=&\prod_{l=1}^Nc(u-\theta_l)c(u+\theta_l)\prod_{\alpha=\e,\e'}(1-2\E^{-\alpha}\sinh(u-\eta))\frac{\sinh(u-6\eta)\cosh(u-\eta)}{\sinh(u-2\eta)\cosh(u-3\eta)},\label{fun-1}\\
\mathbf{d}(u)=&\prod_{l=1}^Nd(u-\theta_l)d(u+\theta_l)\prod_{\alpha=\e,\e'}(1-2\E^{-\alpha}\sinh(u-5\eta))\frac{\sinh u\cosh(u-5\eta)}{\sinh(u-4\eta)\cosh(u-3\eta)},\label{fun-2}\\
\mathbf{b}(u)=& \prod_{l=1}^Nb(u-\theta_l)b(u+\theta_l)\prod_{\alpha=\e,\e'}(1+2\E^{-\alpha}\sinh(u-3\eta))\frac{\sinh
u\sinh(u-6\eta)}{\sinh(u-2\eta)\sinh(u-4\eta)},\label{fun-3}\\
\mathbf{c}(u)=&4^{1-N} c_0\sinh
u\sinh(u-6\eta) \prod_{l=1}^Nc(u-\theta_l)c(u+\theta_l)d(u-\theta_l)d(u+\theta_l),\label{fun-4}
\end{align}
where $c(u)$, $b(u)$ and $d(u)$ are the non-zero elements of the $R$-matrix defined in (\ref{R-element-2}).

The eigenvalue of the transfer matrix $t(u)$, denoted as $\Lambda(u)$, can be parameterized by the following $T$-$Q$ relation \cite{Hao14,ODBA}
\begin{align}
\L(u)=&\,\mathbf{a}(u)\frac{Q_1(u+4\eta)}{Q_2(u)}+\mathbf{d}(u)\frac{Q_2(u-6\eta+\i\pi)}{Q_1(u-2\eta+\i\pi)}+\mathbf{b}(u)
\frac{Q_1(u+2\eta+\i\pi)Q_2(u-4\eta)}{Q_2(u-2\eta+\i\pi)Q_1(u)}\no\\[6pt]
&+\frac{1}{\cosh(u-3\eta)}\left[\frac{\mathbf{c}(u)Q_1(u+2\eta+\i\pi)}{Q_1(u)Q_2(u)}-\frac{\mathbf{c}(-u+6\eta+\i\pi)Q_2(u\hspace{-0.08truecm}-\hspace{-0.08truecm}4\eta)}{Q_1(u-2\eta+\i\pi)Q_2(u-2\eta+\i\pi)}
\right]. \label{T-Q}
\end{align}
The function $Q_i(u)$ in Eq. (\ref{T-Q}) depends on $\bar{N}=4N-2$ parameters $\{\lambda_j|j=1,\ldots,\bar{N}\}$
\bea
 && Q_1(u)=\prod_{k=1}^{\bar N} \sinh\Big(\frac{u-\l_k-2\eta}{2}\Big),\label{Q1} \\[6pt]
&& Q_2(u)=\prod_{k=1}^{\bar N}\sinh\Big(\frac{u+\l_k-2\eta}{2}\Big),\label{Q2}
\eea
and the constant $c_0$ is specified as follows
\bea
&&c_0=-2\E^{-\e-\e'}\lt\{\frac{\cosh(\sigma'-\sigma+2\eta)-\cosh(\bar{N}\eta-\sum_{j=1}^{\bar N}\l_j)}
{\cosh(\frac{\bar{N}\eta}{2}-\frac{1}{2}\sum_{j=1}^{\bar N}\l_j)}\rt\}.\label{c-const}
\eea
The analyticity of $\Lambda(u)$ requires that the residues of $\Lambda(u)$ at $u=\lambda_j+2\eta,\,j=1\ldots,\bar{N}$ must vanish, which leads to the following BAEs
\bea
&&\frac{(1+2\E^{-\e}\sinh(\l_j-\eta))(1+2\E^{-\e'}\sinh(\l_j-\eta))\cosh(\l_j-\eta)}
{4\sinh\l_j\sinh(\l_j-2\eta)}
\no\\
&&=
-\prod_{l=1}^N\sinh\Big(\frac{\l_j-\theta_l-2\eta}{2}\Big)
\sinh\Big(\frac{\l_j+\theta_l-2\eta}{2}\Big)
\cosh\Big(\frac{\l_j-\theta_l}{2}\Big)
\cosh\Big(\frac{\l_j+\theta_l}{2}\Big)\no\\[6pt]
&&\quad\times \frac{c_0\,Q_2(\l_j+\i\pi)}{Q_2(\l_j-2\eta)Q_2(\l_j+2\eta)},\qquad j=1,\ldots,\bar{N}.\label{BAE-M}
\eea

Due to the broken of $U(1)$ symmetry, the $T$-$Q$ relation in Eq. (\ref{T-Q}) includes an inhomogeneous term, which leads to significantly more complex BAEs (\ref{BAE-M}) compared to those (\ref{BAE-p}) in the periodic case. However, under specific conditions, the inhomogeneous term in (\ref{T-Q}) can vanish, as demonstrated in Section \ref{hom;TQ}.
\subsection{\texorpdfstring{Homogeneous $T$-$Q$ relation}{Homogeneous T-Q relation}}\label{hom;TQ}

\paragraph*{Constrained non-diagonal boundaries}

Under the following constraints
\begin{align}
\E^{\sigma'-\sigma}=\E^{-4\mathbf{k}\eta},\quad \mathbf{k}\in \mathbb{Z}, \label{constraint}
\end{align}
the spectrum of the transfer matrix can be parameterized by the following homogeneous $T$-$Q$ relation \cite{Hao14,ODBA}
\begin{align}
\L(u)=&\,\mathbf{a}(u)\frac{Q(u+4\eta)}{Q(u)}+\mathbf{d}(u)\frac{Q(u-6\eta+\i\pi)}{Q(u-2\eta+\i\pi)}+\mathbf{b}(u)
\frac{Q(u+2\eta+\i\pi)Q(u-4\eta)}{Q(u-2\eta+\i\pi)Q(u)},\label{TQ;2}
\end{align}
where the resulting function $Q(u)$ is
\bea
Q(u)=\prod_{j=1}^M\sinh\Big(\frac{u-\l_j-2\eta}{2}\Big) \sinh\Big(\frac{u+\l_j-2\eta}{2}\Big).\label{Q3}
\eea
Here $M$ is a non-negative integer and takes the following values
\begin{align}
M=\begin{cases}
N-\mathbf{k}, & \mathbf{k}\leq -N, \\
N+\mathbf{k}+1, & \mathbf{k}\geq N+1,\\
N-\mathbf{k}, & 1-N\leq \mathbf{k}\leq N,\\
N+\mathbf{k}-1, & 1-N\leq \mathbf{k}\leq N.\\
\end{cases}\label{int;M}
\end{align}
The resulting BAEs now read
\bea
&&\hspace{-1.2truecm}\prod_{l=1}^N\frac{\sinh\Big(\frac{\l_j-\theta_l-2\eta}{2}\Big)\sinh\Big(\frac{\l_j+\theta_l-2\eta}{2}\Big)}
{\sinh\Big(\frac{\l_j-\theta_l+2\eta}{2}\Big)\sinh\Big(\frac{\l_j+\theta_l+2\eta}{2}\Big)}\,\prod_{\alpha=\e,\e'}
\frac{(1-2\E^{-\alpha}\sinh(\l_j+\eta))}
{(1+2\E^{-\alpha}\sinh(\l_j-\eta))}\no\\[6pt]
&&\hspace{-1.2truecm} =-\frac{\sinh(\l_j+2\eta)\cosh(\l_j-\eta)}{\sinh(\l_j-2\eta)\cosh(\l_j+\eta)}
\frac{Q(\l_j-2\eta)Q(\l_j+4\eta+\i\pi)}{Q(\l_j+6\eta)Q(\l_j+\i\pi)},\quad j=1,\ldots,M.\label{BAE-2}
\eea

Although the $U(1)$ symmetry remains broken, the system exhibits an additional symmetry under Eq. (\ref{constraint}), which ensures the existence of a homogeneous $T$-$Q$ relation. In this case, the integer $M$ is fixed by the system parameters. One can then find a proper ``local vacuum" to proceed with the generalized Bethe ansatz approach and study the physical quantities of the model \cite{Cao03,Zhang21}. When $M \geq 2N$, the $T$-$Q$ relation (\ref{TQ;2}) with $M$ Bethe roots may provide the complete set of eigenvalues of the transfer matrix. When $0 \leq M < 2N$, two $T$-$Q$ relations are required to parameterize the full spectrum of the transfer matrix, with the number of Bethe roots being $M$ and $2N-M-1$, respectively.
Such degenerate points exist in various integrable models, e.g., the anisotropic spin-$\frac{1}{2}$ chains with non-diagonal boundary fields \cite{Cao03,Nepomechie03a,Nepomechie03b,Cao1d,Zhang21}).

\vspace{0.2cm}

\paragraph{Diagonal boundaries} As the boundary parameter $\epsilon$ approaches infinity $\epsilon \to +\infty$, the resulting $K$-matrices become diagonal
\bea
K^-(u)={\id},\quad K^+(u)=\cal M,\label{Diagonal-Kmatrix}
\eea
where the matrix $\cal M$ is defined in (\ref{M-matrix}).
In this case, the $K$-matrices automatically satisfy the operator relation
\bea
\lim_{\eta\rightarrow0}\{K^+(u)K^-(u)\}={\id}.\label{Int-K}
\eea
The $U(1)$-symmetry of the IK model is now recovered, and one can also use homogeneous $T$-$Q$ relations to parameterize the spectrum of the transfer matrix.
The functions $\mathbf{a}(u)$, $\mathbf{b}(u)$ and $\mathbf{d}(u)$ given by Eqs. (\ref{fun-1})-(\ref{fun-3}) reduce to \cite{Mez92}
\begin{align}
\bar{\mathbf{a}}(u)&=\prod_{l=1}^Nc(u-\theta_l)c(u+\theta_l)\,\frac{\sinh(u-6\eta)\cosh(u-\eta)}{\sinh(u-2\eta)\cosh(u-3\eta)},\\
\bar{\mathbf{d}}(u)&=\prod_{l=1}^Nd(u-\theta_l)d(u+\theta_l)\,\frac{\sinh u\cosh(u-5\eta)}{\sinh(u-4\eta)\cosh(u-3\eta)},\\
\bar{\mathbf{b}}(u)&=\prod_{l=1}^Nb(u-\theta_l)b(u+\theta_l)\,\frac{\sinh u\sinh(u-6\eta)}{\sinh(u-2\eta)\sinh(u-4\eta)}.
\end{align}
The $T$-$Q$ relation (\ref{T-Q}) now can be simplified as the following one \cite{Mez92}
\begin{align}
\L(u)=\bar{\mathbf{a}}(u)\frac{Q(u+4\eta)}{Q(u)}+\bar{\mathbf{d}}(u)\frac{Q(u-6\eta+\i\pi)}{Q(u-2\eta+\i\pi)}+\bar{\mathbf{b}}(u)\frac{Q(u-4\eta)Q(u+2\eta+\i\pi)}{Q(u-2\eta+\i\pi)Q(u)},
\label{T-Q-2}
\end{align}
where the function $Q(u)$ is defined as
\bea
Q(u)=\prod_{j=1}^M\sinh\Big(\frac{u-\l_j-2\eta}{2}\Big) \sinh\Big(\frac{u+\l_j-2\eta}{2}\Big).\label{Q3-1}
\eea
In this case, the integer $M$ is adjustable and can take the following values
\bea
M=0,1,\ldots,N.\label{M-value-4}
\eea
The resulting BAEs for the Bethe roots $\{\l_j\}$ in (\ref{T-Q-2}) are
\bea
&&\prod_{l=1}^N\frac{\sinh\Big(\frac{\l_j-\theta_l-2\eta}{2}\Big)\sinh\Big(\frac{\l_j+\theta_l-2\eta}{2}\Big)}
{\sinh\Big(\frac{\l_j-\theta_l+2\eta}{2}\Big)\sinh\Big(\frac{\l_j+\theta_l+2\eta}{2}\Big)}
\,\frac{\sinh(\l_j-2\eta)\cosh(\l_j+\eta)}{\sinh(\l_j+2\eta)\cosh(\l_j-\eta)}\no\\[6pt]
&&=-\frac{Q(\l_j-2\eta)Q(\l_j+4\eta+\i\pi)}{Q(\l_j+6\eta)Q(\l_j+\i\pi)},
\qquad j=1,\ldots,M.\label{BAE-3}
\eea
In the following sections, we will construct integrable IK Gaudin models with open boundaries and derive their exact spectra.
\section{IK Gaudin model with open boundaries}
\setcounter{equation}{0}
Following the approach outlined in Section \ref{Sec;Gaudin;Periodic}, one can construct the associated Gaudin operators $\{H_j\}$.
We expand the transfer matrix $t(\theta_j)$ around $\eta=0$, specially as follows
\begin{align}
&t(\theta_j)=\kappa\,\mathsf{t}_0(\theta_j)(\id+\eta H_j+\cdots), \quad j=1,\dots, N,\no\\[6pt]
&H_j=\lt.\frac{\partial \ln (t(\theta_j)/\kappa)}{\partial\eta}\rt|_{\eta=0}.\label{exp-t}
\end{align}
Equation (\ref{cond-1}) implies that
\begin{align}
\mathsf{t}_0(\theta_j)&=\lim_{\eta\rightarrow 0} \tr_0 \lt\{
\prod_{l\neq j}^{N}\sinh(\theta_j-\theta_l)\prod_{l=1}^{N}\sinh(\theta_j+\theta_l)K^+_0(\theta_j)P_{0,j}K^-_0(\theta_j)\rt\}\no\\
&=\prod_{l\neq j}^{N}\sinh(\theta_j-\theta_l)\prod_{l=1}^{N}\sinh(\theta_j+\theta_l)\,\lim_{\eta\rightarrow0}\{K^-_j(\theta_j)K^+_j(\theta_j)\}.
\end{align}
To ensure that the resulting Gaudin operators form a commuting family, i.e.,
\begin{align*}
[H_i,\,H_j]=0,\qquad i,j=1,2,\dots,N,
\end{align*}
which is essential for the integrability of the corresponding Gaudin model \cite{Gau76}, we require that $\mathsf{t}_0(\theta_j)$ be proportional to the identity operator (see Section \ref{Sec;Gaudin;Periodic})
\begin{align}
\lim_{\eta\rightarrow0}\left\{K^-_j(\theta_j)K^+_j(\theta_j)\right\}\propto {\id}\quad .\label{cons-open}
\end{align}
Equation (\ref{cons-open}) gives rise to the following restrictions for the boundary parameters
\begin{align}
\lim_{\eta\to 0}\E^{\sigma}=\E^{\sigma'},\quad \lim_{\eta\to 0}\E^{\epsilon'}=-\E^\epsilon.\label{cons-2}
\end{align}
Without loss of generality, we assume that the boundary parameters $\epsilon$, $\epsilon^{\prime}$ and $\sigma$ are independent of the crossing parameter $\eta$, while $\sigma^{\prime}$ depends on $\eta$. From these assumptions, we get the following identities
\bea
\sigma^{\prime}=\sigma+\bar\sigma\eta,\quad \epsilon'=\epsilon+\i \pi.\label{cons-1}
\eea
As a consequence, the following equation can be obtained
\begin{align}
\mathsf{t}_0(\theta_j)=\prod_{l\neq j}^{N}\sinh(\theta_j-\theta_l)\prod_{l=1}^{N}\sinh(\theta_j+\theta_l)\,w(\theta_j)\times{\id},\quad w(\theta_j)=(1-4\E^{-2\epsilon}\sinh^2\theta_j). \label{t0;theta}
\end{align}

Using the initial condition of $R$-matrix (\ref{Int-R}) and the QYBE (\ref{QYB}), the double row transfer matrix at the point $u=\theta_j$ can be expressed as \cite{ODBA}
\begin{align}
t(\theta_j)=&\,\kappa R_{j,j-1}(\theta_j-\theta_{j-1})\cdots R_{j,1}(\theta_j-\theta_1)K_j^-(\theta_j)R_{1,j}(\theta_j+\theta_1)\cdots R_{j-1,j}(\theta_j+\theta_{j-1})\no\\
&\times R_{j+1,j}(\theta_j+\theta_{j+1})\cdots R_{N,j}(\theta_j+\theta_N)\tr_0\{K_0^+(\theta_j)P_{0,j}R_{j,0}(2\theta_j)\}\no\\
&\times R_{j,N}(\theta_j-\theta_N)\cdots R_{j,j+1}(\theta_j-\theta_{j+1}).
\end{align}
Then, the following Gaudin operator can be constructed
\begin{align}
H_j=\Gamma_j(\theta_j)+ \sum_{l\neq j }^N\bar\Gamma_{j,l}(\theta_j,\theta_l),\label{off-gaudin-H}
\end{align}
where the operator $\Gamma_j(\theta_j)$ and $\bar\Gamma_{j,l}(\theta_j,\theta_l)$ are defined as
\begin{align}
\Gamma_j(\theta_j)&=\frac{1}{\sinh(2\theta_j)w(\theta_j)}
\lim_{\eta\to0}\frac{\partial}{\partial\eta}\Big[K_j^-(\theta_j)\,\tr_0\{K_0^+(\theta_j)P_{0,j}R_{j,0}(2\theta_j)\}\Big],\\[6pt]
\bar\Gamma_{j,l}(\theta_j,\theta_l)&=\frac{r_{j,l}(\theta_j-\theta_l)}{\sinh(\theta_j-\theta_l)} +\left.\frac{K_j^-(\theta_j) r_{l,j}(\theta_j+\theta_l)K_{j}^{+}(\theta_j)}{w(\theta_j)\sinh(\theta_j+\theta_l)}\right|_{\eta\rightarrow 0},
\end{align}
and the operator $r_{j,l}(u)$ is given by (\ref{ope-r}) and (\ref{operator;r}). Here $\Gamma_j(\theta_j)$ describes the on-site potential, while $\bar\Gamma_{j,l}(\theta_j, \theta_l)$ represents a site-dependent, long-range two-site interaction.
One can also use the spin-1 operators to expand the Gaudin operator. After some tedious calculations, we get the expression of $\Gamma_j(u)$
\begin{align}
\Gamma_j(u)=&
-\frac{\E^{-\sigma } \sinh u \left(4 \E^u-\E^{\epsilon } \left(\bar{\sigma }+2\right)\right)}{\E^{2\epsilon }-4 \sinh ^2u}(S_j^-)^2-\frac{\E^{\sigma } \sinh u \left(4 \E^{-u}+\E^{\epsilon} \left(\bar{\sigma }+2\right)\right)}{\E^{2\epsilon}-4 \sinh ^2u}(S_j^+)^2\no\\
&-\frac{4\sinh(2u)}{\E^{2\epsilon}-4\sinh ^2u}(S_j^z)^2-\frac{4 \sinh u \left(\E^{\epsilon }-\left(\bar{\sigma }+2\right) \sinh u\right)}{\E^{2\epsilon}-4 \sinh ^2u}S_j^z\no\\
&-\frac{\E^{2\epsilon} (5+7\cosh (2u))+5+4 \cosh (2 u)-9\cosh(4u)}{\sinh (2u) \left(\E^{2\epsilon}-4 \sinh ^2u\right)}\,\id_j.
\end{align}
The two-site interaction $\bar\Gamma_{j,l}(u)$ comprises not only the operator $r_{i,j}(u)$ given in Eq. (\ref{operator;r}), but also two additional operators on site $j$
\begin{align}
K_j^-(u)|_{\eta\to0}=&\,\E^{-\sigma -\epsilon}\sinh u\, (S_j^-)^2 +\E^{\sigma -\epsilon }\sinh u\, (S_j^+)^2 \no\\
&+2\E^{-\epsilon } \sinh u\,(S_j^z)^2+\left(1-2 \E^{-\epsilon } \sinh u\right)\id_j,\\
K_j^+(u)|_{\eta\to0}=&-\E^{-\sigma -\epsilon}\sinh u\, (S_j^-)^2 -\E^{\sigma -\epsilon }\sinh u\, (S_j^+)^2 \no\\
&-2\E^{-\epsilon } \sinh u\,(S_j^z)^2+\left(1+2 \E^{-\epsilon } \sinh u\right)\id_j.
\end{align}

Unlike the two-site interaction $\Gamma_{j,l}(\theta_j, \theta_l)$ in the periodic system (see Eq. (\ref{new-1})), $\bar\Gamma_{j,l}(\theta_j,\theta_l)$ in the open system depends on the inhomogeneous parameters $\theta_{j,l}$ and the boundary parameters $\sigma$ and $\epsilon$.

The Gaudin operator defined in Eq. (\ref{off-gaudin-H}) is exactly solvable.
Equation (\ref{exp-t}) allows us to obtain the eigenvalues of the Gaudin operators from the exact spectrum of the transfer matrix $t(u)$ at $u=\theta_j$ as follows
\begin{align}
E_j=\lt.\frac{\partial\ln (\Lambda(\theta_j)/\kappa)}{\partial\eta}\rt|_{\eta=0}.
\end{align}

As demonstrated in Section \ref{TQ;Open}, the homogeneous $T$-$Q$ relation exhibits a significantly simpler form compared to its inhomogeneous counterpart, yielding BAEs that are more tractable for analytical and numerical treatment. Consequently, in the following section, we will consider the exact solutions of the IK Gaudin model with boundary conditions specified by Eqs. (\ref{constraint}) and (\ref{Diagonal-Kmatrix}), respectively.

\section{Exact solution of the IK Gaudin model with open boundaries}\label{BA-IK-Open}

\setcounter{equation}{0}
\setcounter{remark}{0}

\subsection{Constrained non-diagonal boundaries}
%%%%%%%%%%%%%%%%%%
%\label{EMT} \setcounter{equation}{0}

Following Eqs. (\ref{constraint}) and (\ref{cons-2}), one can construct the corresponding Gaudin operator (\ref{off-gaudin-H}) by letting $\bar\sigma=-4\mathbf{k}$ and $\E^{\epsilon'}=-\E^{\epsilon}$. It should be noted that the extra constraint (\ref{constraint}) only affects the specific expression of the operator $H_j$, without altering its underlying structure (the position of non-zero entries in the matrix).

The eigenvalues of the IK Gaudin operators read
\begin{align}
E_j=&\lt.\frac{\partial\ln (\Lambda(\theta_j)/\kappa)}{\partial\eta}\rt|_{\eta=0}\no\\
=&\sum_{l=1}^M\frac{4\sinh\theta_j}{\cosh\theta_j-\tilde\mu_l}-6\coth \theta_j-\tanh\theta_j+\frac{4\E^{-2\epsilon}\sinh(2\theta_j)}{1-4\E^{-2\epsilon}\sinh^2(2\theta_j)}\no\\
&+\sum_{k \neq j}^N\lt[\frac{1-5\cosh(\theta_j-\theta_k)}{\sinh(\theta_j-\theta_k)}+\frac{1-5 \cosh(\theta_j+\theta_k)}{\sinh(\theta_j+\theta_k)}\rt],\label{homo-1}
\end{align}
and the corresponding BAEs are
\begin{align}
&-\sum_{k \neq j}^M \frac{\tilde\mu_j+3\tilde\mu_k}{\tilde\mu_j^2-\tilde\mu_k^2}+\sum_{l=1}^N \frac{1}{\tilde\mu_j-\cosh\theta_l}+\frac{4\tilde\mu_j}{\E^{2\epsilon}+4-4{\tilde\mu}_j^2}=0,\qquad j=1,\ldots,M.\label{homo-2}
\end{align}

Notably, there is a one-to-one correspondence between  $\tilde\mu_j$ in Eq. (\ref{homo-2}) and $\l_j$ in Eq. (\ref{BAE-2}): $\tilde\mu_j=\lim\limits_{\eta\to0}\cosh\lambda_j$. From Eqs. (\ref{homo-1}) and (\ref{homo-2}), we observe that the eigenvalue of the Gaudin operator $H_j$ depend on the set $\{\tilde\mu_j\}$ and $\epsilon$, but is independent of $\sigma$ and $\bar\sigma$.
The numerical solutions of Eq. (\ref{homo-2}) for small-scale systems are presented in Table \ref{tab2}.
\begin{table}[htbp]
\centering
\begin{tabular}{|c|c|c|r|c|}
\hline
 $\tilde\mu_1$ & $\tilde\mu_2$ & $\tilde\mu_3$ &$E_1$~~~~ &$d$ \\
 \hline
 1.0462 & $-$31.3108 & 0.0541 & $-$17.5222 & 1\\
 0.1573 & 1.0461 &-- & $-$17.5011 & 2\\
 1.3552 & 1.0504 & --& $-$16.3699 & 2\\
 $-$1.5098 & 1.0453 &-- & $-$15.3492 & 2\\
 1.0478 & 1.3151 & $-$1.2458 & $-$11.8797 & 1 \\
 $-$1.1210 & 1.0471 & 1.1846 & $-$1.9845 & 1 \\
 0.2520$-$0.7199$\i$ & 0.2520+0.7199$\i$ &-- &  28.9411 & 2\\
 $-$8.7360 & $-$1.6847 & -- &30.4393  & 2\\
 0.8132$-$0.3340$\i$  & 0.8132+0.3340$\i$ & 1.3385 & 32.7812 &1 \\
 $-$0.2126 & 1.3237 & 14.4331 & 36.8243  & 1\\
 $-$0.4131 & 1.3235 &--  & 36.8791 & 2\\
 2.0772 & 1.3188 &-- & 39.7611 & 2\\
 27.7924 & $-$0.0643 & 1.1818 & 46.1381  &1 \\
 $-$0.1616 & 1.1820 &-- & 46.1579 & 2\\
 $-$1.4080 & 1.1716  &-- & 48.6891 & 2\\
 1.5999 & 1.1923 &-- & 49.1452& 2 \\
 1.2134$-$0.0445$\i$ & 1.2134+0.0445$\i$ & $-$1.3308 & 52.8331 &1 \\
 \hline
\end{tabular}
\caption{Numeric solutions of Eq. (\ref{homo-2}) with $N=3$, $\{\theta_1,\theta_2,\theta_3\}=\{-0.40,0.18,0.67\}$ and $\{\epsilon,\sigma, \bar{\sigma }\}=\{0.50,0.12,-4\}$. From Eq. (\ref{int;M}), the number of Bethe root $M$ can be 2 or 3. The eigenvalue of $H_1$ given by Eq. (\ref{homo-1}) matches the exact diagonalization results. Here $d$ represents the degeneracy of the energy level.}\label{tab2}
\end{table}
\subsection{Diagonal open boundaries}
%\label{EMT} \setcounter{equation}{0}

When $\epsilon\to+\infty$, the expression of the Gaudin operator (\ref{off-gaudin-H})  can be simplified as follows
\begin{align}
\Gamma_j(\theta_j)&=(-6 \coth\theta_j - \tanh\theta_j)\times \id,\\
\Gamma'_{j,l}(\theta_j,\theta_l)&=\frac{r_{j,l}(\theta_j-\theta_l)}{\sinh(\theta_j-\theta_l)} +\frac{ r_{l,j}(\theta_j+\theta_l)}{\sinh(\theta_j+\theta_l)}.
\end{align}

The corresponding eigenvalues of the IK Gaudin operators in terms of the Bethe roots are
\begin{align}
E_j=&\sum_{k=1}^M \frac{4\sinh \theta_j}{\cosh\theta_j-\bar\mu_k}-\tanh\theta_j-6\coth\theta_j\no\\
&+\sum_{k \neq j}^N\lt[\frac{1-5\cosh(\theta_j-\theta_k)}{\sinh(\theta_j-\theta_k)}+\frac{1-5 \cosh(\theta_j+\theta_k)}{\sinh(\theta_j+\theta_k)}\rt],\label{diag-2}
\end{align}
where $\{\bar\mu_j|j=1,\ldots,M\}$ satisfy the following BAEs
\begin{align}
&\sum_{l=1}^N \frac{1}{\bar\mu_j-\cosh\theta_l}-\sum_{k \neq j}^M \frac{\bar\mu_j+3\bar\mu_k}{\bar\mu_j^2-\bar\mu_k^2}=0,\qquad j=1,\dots, M.\label{homo-3}
\end{align}
Analogously, $\bar\mu_j$ in Eq. (\ref{homo-3}) and $\l_j$ in Eq. (\ref{BAE-3}) has a one-to-one correspondence $\bar\mu_j=\lim\limits_{\eta\to0}\cosh\lambda_j$. The numerical solutions of Eq. (\ref{homo-3}) for small-scale systems are presented in Table \ref{tab3}.
\begin{table}[htbp]
\centering
\begin{tabular}{|c|c|c|r|c|}
\hline
 $\bar\mu_1$ & $\bar\mu_2$ & $\bar\mu_3$ &$E_1$~~~~ &$d$ \\
 \hline
1.5573 & -- & -- & $-$44.7405 & 5\\
$-$0.3537 & 1.5480 &-- & $-$43.6642 & 3\\
-- & -- & -- & $-$41.2908 & 7 \\
0.2509$-$0.6787$\i$ & 0.2509+0.6787$\i$ &-- & $-$38.9182 & 3\\
1.5348 & 1.0476 & $-$1.2797 & 4.8930 &1\\
1.0480 & -- & -- & 8.3428  & 5\\
0.3834 & 1.0468 & --& 8.9485 & 3 \\
\hline
\end{tabular}
\caption{Numeric solutions of Eq. (\ref{homo-3}) with $N=3$, $\{\theta_1,\theta_2,\theta_3\}=\{0.40,0.18,1.20\}$. The eigenvalue of $H_1$ given by Eq. (\ref{homo-3}) matches the exact diagonalization results. Here $d$ represents the degeneracy of the energy level.}\label{tab3}
\end{table}

\section{Conclusions}
\label{EMT} \setcounter{equation}{0}

In this paper, we study the integrable IK Gaudin model with both periodic and open boundary conditions. We construct the Gaudin operator by expanding the inhomogeneous transfer matrix at the point $u=\theta_j$, which ensures the solvability of the Gaudin operator. By leveraging the exact solutions of the IK model, obtained through the Bethe-ansatz method, we finally get the eigenvalue spectrum of the Gaudin operator in terms of the Bethe roots, which are determined by the corresponding Bethe ansatz equations (BAEs). Numerical computations have also been done to validate our analytical results.

The IK model with open boundary conditions deserves further elaboration. In this paper, we only consider the non-diagonal $K$-matrices in Eqs. (\ref{Kf}) and (\ref{Kz}). It should be emphasized that our method remains applicable to other $K$-matrix classes examined in Refs. \cite{Lima99,Nep2002}. Since the Gaudin operator's form explicitly depends on boundary parameters, distinct $K$-matrix configurations will result in physically distinct Gaudin operators.

Future work may involve further analysis of BAEs. It will be interesting to explore the existence of infinite Bethe roots or singular physical solutions of the corresponding BAEs, which may be essential for the completeness of our BAEs and the explanation of the degeneracy.

Another open question is the construction of eigenstates for the Gaudin operator. When the system retains $U(1)$ symmetry, the eigenstates can be constructed using the algebraic Bethe ansatz method. However, this approach requires further adaptation or generalization to address the case withou $U(1)$ symmetry.

\section*{Acknowledgments}
We thank Professors Vladimir Korepin and Kun Hao for valuable discussions.
Financial support from the National Natural Science Foundation of China (Grant Nos. 12105221, 12205235, 12247103, 12074410, 12047502), the Strategic Priority Research Program of the Chinese Academy of Sciences (Grant No. XDB33000000), Shaanxi Fundamental Science Research Project for Mathematics and Physics (Grant Nos. 22JSZ005),  the Scientific Research Program Funded by Shaanxi Provincial Education Department (Grant No. 21JK0946), Beijing National Laboratory for Condensed Matter Physics (Grant No. 202162100001), and the Double First-Class University Construction Project of Northwest University is gratefully acknowledged.

\end{document}